\documentclass[a4paper]{article}

\usepackage[dvips]{graphicx}
\usepackage{bm}

\newcommand{\rng}[1]{\mathring{#1}}
\newcommand{\ns}{\!\!\!\!\!}
\newcommand{\tl}[1]{\tilde{#1}}

\begin{document}

\twocolumn
[
\title{\textsc{Gyroscope deviation from geodesic motion:\\
               quasiresonant oscillations on a circular orbit}}
\author{O. B. Karpov\\
        \textit{Moscow State Mining University}\\
        \textit{119991, Moscow, Russia}}
\date{\today}
\maketitle \vspace{-11ex}

\renewcommand{\abstractname}{ }\abstractname
   \begin{abstract}
{\bf {\sf General relativistic spin-orbit interaction gives rise
to quasiresonant oscillation of the gyroscope mass center along
the orbital normal. The oscillation amplitude appears to be
measurable by the present-day instruments. The influence of
oblateness of the field source has been investigated.}}
  \end{abstract}\\

\vspace{6.0ex}%
]

\section{Introduction}

    In general relativity a motion of a spinning test body (gyroscope)
undergoes a spin-orbit interaction in two aspects: 1) an influence
of the orbital motion on the gyroscope rotation axes orientation,
2) an influence of the gyroscope intrinsic momentum (spin) on its
orbit. The first one is comparative simple when the spin parallel
transport is assumed. It is admissible if a deviation from a
geodesic motion is small. The Fermi-Walker transport along
appointed world line is not complicated too.  The parallel
transport in an spherically symmetric field along a geodesic leads
to a precession of the gyroscope axes known as the geodetic or de
Sitter precession \cite{1}. In the field of a rotating mass the
gyroscope axes undergoes the Schiff precession \cite{2} being to
verify in the Gravity Probe B experiment (see \cite{3} for
details).

    In this work the second aspect of the spin-orbit interaction
is considered.  The orbital motion of the gyroscope is a
sophisticated problem that has not been resolved in full till now
even in post-Newtonian approximation. There do exist several
various approaches with different results in the main
approximation (for example \cite{4,5,6,7,8,9,10}). The only
covariant general relativistic equations of motion of the spinning
test particles are well-known Papapetrou equations \cite{5}. This
equations set is incomplete and requires supplementary conditions.
It is generally accepted that these conditions single out the
representative point as a gyroscope mass center, but there exist
diverse other opinions \cite{9,10,11,12}. Besides, the Papapetrou
equations or alternative ones are very complicated. Its
investigation is limited usually by a general analysis and an
examination of the effects is restricted as a rule by the motion
of the gyroscope with a vertical spin, i.e. the gyroscope axes
orthogonal the orbital plane. It is known for example that such
gyroscope moves on a circular orbit with a velocity differing from
one of a body without spin \cite{13}. A gyroscope with a
horizontal spin goes out the geodesic plane. A quasiresonant
character of the spin-orbit interaction in this case has been
revealed first in works \cite{14, 15}.

    In the present work the motion of the gyroscope with the
horizontal spin is investigated and the general relativistic
effect of quasiresonant beating is proposed. Due to a small
denominator the speed of light in an oscillation amplitude is
cancelled and  therefore the effect turns out quite sizable. The
obvious physical interpretation of the effect is given. This
effect does not depend on supplementary conditions and is the same
in the different approaches \cite{4,5,6,7,8,9,10}. A significant
simplification of the description is achieved by the expansion of
the equations of motion linear in the displacement from a
geodesic. Instead of studying the intricate gyroscope orbit having
not described before, the small oscillation is investigated. This
oscillation gives sufficient information about the gyroscope
orbit. It is shown that a Newtonian nonsphericity of the field
source causes a specific effacing of the quasiresonant beating,
retaining the oscillation amplitude measurable.

   In the following calculations orthonormal bases are used, Greek
indices run from 0 to 3, Latin indices - from 1 to 3. Signature is
$(-++\:+)$.

      \section{The essence of the effect}

     General relativistic spin-orbit acceleration $a$
deviating a gyroscope mass center from a geodesic is of order of
   \begin{equation}\label{eq:1}
                a \sim \epsilon\,\frac{S}{\lambda}\,g,
   \end{equation}
where $\epsilon = GM/c^{2}r$ is the relativistic small parameter,
$g = GM/r^{2}$ is Newtonian acceleration due to gravity, $M$ is
the source mass, $S$ is the the spin of the gyroscope, $\lambda$
is its orbital momentum, $c$ is the speed of light, $G$ is the
gravitational constant. The motion of the rotating body mass
center essentially (in the main approximation (\ref{eq:1}))
depends on the reference frame in which it has been obtained.
General expression of the spin-orbit acceleration in the main
post-Newtonian approximation (\ref{eq:1}) is \cite{7, 16}
   \begin{eqnarray}\label{eq:2}
 \bm{a} = 3 \frac{GM}{mc^{2}r^{3}}[\bm{S}\times\bm{v}
    &\ns+\ns&(2 - \sigma)\,\bm{\hat{r}}\,
 (\bm{S}(\bm{\hat{r}}\times\bm{v})) \nonumber\\
    &\ns-\ns&
(1 + \sigma)(\bm{v}\bm{\hat{r}})(\bm{S}\times\bm{\hat{r}})].
   \end{eqnarray}
Parameter $\sigma$ numbers the different mass centers: $\sigma =
0$ corresponds to the Dixon \cite{6} and Pirani \cite{17}
conditions (the intrinsic mass center),  $\sigma = 1$ corresponds
to the Corinaldesi-Papapetrou conditions \cite{18} (the mass
center defined in the "rest" frame in which the gyroscope moves
with a velocity $\bm{v}$), and $\sigma = 1/2$ leads to the results
of Fock \cite{4} and \cite{9, 10}. For a circular orbit ($\bm{vr}
= 0$) of the gyroscope with its axis lying in the orbital plane
($\bm{S}(\bm{r} \times\bm{v}) = 0$), independently of the
parameter $\sigma$, the spin-orbit acceleration (\ref{eq:2}) is
   \begin{equation}\label{eq:3}
   \bm{a} = 3\frac{GM}{mc^{2}r^{3}}\,\bm{S}\times\bm{v}.
   \end{equation}
Parallel transport of the spin vector $\mathbf{S}$ means that in
process of revolution the acceleration (\ref{eq:3}) is directed
along the orbital normal $\mathbf{e}_{3}$ and is periodic in time
$\tau$,
   \[
       \bm{a} = \bm{e}_{3}\,\epsilon\,\frac{S}{\lambda}\,g
       \cos(\omega_{s} \tau + \beta).
   \]
The frequency $\omega_{s}$ differs from the orbital frequency
$\omega$ because of geodetic precession $\Omega^{G}$,
   \begin{equation}\label{eq:4}
       \Delta\omega = \omega - \omega_{s} = \Omega^{G} =
                        \frac{3}{2}\,\epsilon\,\omega.
   \end{equation}
On the other hand, the frequency of free tidal oscillation along
the orbital normal is equal to the orbital frequency. It leads to
an almost resonant beating with the modulation frequency
(\ref{eq:4}) and maximum amplitude
   \begin{equation}\label{eq:5}
   A = \frac{a}{2\omega\,\Delta\omega} =  \frac{S}{\lambda}\,r.
   \end{equation}
Note the cancellation of the speed of light $c$ in the amplitude
(\ref{eq:5}) by the relativistic small denominator $\Delta\omega$
(\ref{eq:4}). During the time $\tau \ll (\Omega^{G})^{-1}$
quasiresonant oscillation enhances linearly at a rate
   \begin{equation}\label{eq:6}
   A\,\Delta \omega = \frac{3}{2}\,\epsilon\,\frac{S}{\lambda}\,v
   \end{equation}
and reaches the values measurable with the present-day
instruments. For example, in the case of a gyroscope of $10^{-1}$
m in dimension with $10^{-1}$ s period of intrinsic rotation in a
near-Earth orbit $r \simeq 7\cdot10^{3}$ km, we get the following
values
   \begin{equation}\label{eq:7}
   \epsilon \sim 10^{-10},\;\; \frac{S}{\lambda} \sim 10^{-9},\;\;
              A\,\Delta \omega \sim 10^{-9}\mbox{ cm/day}.
   \end{equation}
Parasitic effects of nonrelativistic origin are mutually cancelled
in the symmetric relative oscillations of two gyroscopes with
antiparallel spins.

     \section{The calculation of the net\\ effect}

   Static, spherically symmetric gravitational field in the post-
Newtonian approximation is described by the tetrad
   \begin{equation}\label{eq:8}
     \rng{\bm{e}}^{\mu} = \{(1 - \epsilon)\,cdt,\;
     (1 + \epsilon)\,dr,\; r\sin\theta\,d\phi,\; -r\,d\theta\}
   \end{equation}
which represents the rest observers in the Schwarz\-schild metric.
``Electric'' $E$ and ``magnetic'' $B$ parts of the Riemann tensor
$\mathcal{R}$ (see, for example, \cite{8, 16})
 \[
     E_{ij}=\mathcal{R}_{i0j0}, \quad
    2B_{ij}=\mathcal{R}_{iomn}\varepsilon^{mn}\,_{j}
 \]
 in this frame are
   \begin{eqnarray}\label{eq:9}
    \rng{E}_{ij}&\ns=\ns& n^{2}\mathrm{diag}\{-2,\; 1,\; 1\},\quad
    \rng{B}_{ij} = 0, \\
                &\ns \ns& n^{2} = GM/r^{3}.\nonumber
   \end{eqnarray}
Transition to the orbital frame $\bm{e}^{\nu}$ is fulfilled by the
boost $\rng{\bm{e}}^{\mu} = L^{\mu}\,_{\nu}\, \bm{e}^{\nu}$ in
$\rng{\bm{e}}_{2}$-direction. The Lorentz matrix $L$ has the
standard form. Namely, the components of the 4-velocity of the
fiducial orbital motion $\phi = nt = \omega \tau$ are
   \begin{equation}\label{eq:10}
       u^{\mu} = L^{\mu}\,_{0} = \gamma\{1,\; 0,\;\beta,\;0\},
     \vspace{-3ex}
   \end{equation}
where
  $$
  \begin{array}{rclrcl}
\gamma&\ns=\ns&(1 - \beta^{2})^{-1/2},\;& \beta&\ns=\ns&v/c, \\
     v&\ns=\ns&(1 + \epsilon)nr,        &\omega&\ns=\ns&\gamma v/r
      = n(1 + 3\epsilon/2),
  \end{array}
  $$
and $\tau$ is the proper time. The $\bm{e}_{1}$ axis is directed
along the current radius-vector, the $\bm{e}_{2}$ axis is along
the orbital motion velocity, and $\bm{e}_{3}$ is orthogonal to the
orbital plane:
   \[
   L^{1}\,_{1} = 1,\quad L^{2}\,_{2} = \gamma,\quad L^{3}\,_{3} = 1.
   \]
The angular velocity vector $\bm{\Omega}$ of rotation of the
orbital triad $\nabla_{u}\bm{e}^{i} = \Omega^{i}\,_{k}\bm{e}^{k}$
has the only component
   \begin{equation}\label{eq:11}
 \Omega_{3} = \Omega_{12} \stackrel{\rm def}{=}\omega_{s} = n.
   \end{equation}
The transformation of the ``magnetic'' matrix \cite{16}
   \begin{eqnarray}\label{eq:12}
B_{ij}&\ns=\ns&4\rng{B}_{kl}L^{[k}\,_{i}u^{0]}L^{[l}\,_{j}u^{0]}
                                                  \nonumber\\
      &\ns \ns& \mbox{} -
 \rng{B}_{pq}\varepsilon^{p}\,_{km}\varepsilon^{q}\,_{ln}
             L^{k}\,_{i} u^{m} L^{l}\,_{j} u^{n}  \nonumber\\
      &\ns \ns& \mbox{} -
 4\rng{E}_{km}\varepsilon^{m}\,_{ln}
 L^{[k}\,_{(i}u^{0]}L^{l}\,_{j)}u^{n}
   \end{eqnarray}
leads to an appearance in orbital frame of the component
   \begin{equation}\label{eq:13}
      B_{31} = \beta(\rng{E}_{33} - \rng{E}_{11}) =
               3n^{2}\beta.
   \end{equation}
The transformation of the ``electric'' matrix is analogous to
(\ref{eq:12}) with a substitution $B \rightarrow E,\, E
\rightarrow -B$ (see \cite{16}). The result is
\[
 E_{11}=-2n^{2}(1 + \epsilon/2),\;\; E_{22}=n^{2},
\]
   \begin{equation}\label{eq:14}
 E_{33}=n^{2}(1 + 3\epsilon)=\omega^{2}.
   \end{equation}
Note the invariance of the component $E_{22}$ parallel to the
boost and the equality $E_{33} = \omega^{2}$ being exact.

   The equation of motion of the gyroscope mass center in the
orbital frame is the equation of geodesic deviation with the
spin-orbit acceleration (\ref{eq:2}) in its right-hand
side\footnote {The equation (\ref{eq:15}) can be obtained by an
expansion of the Papapetrou equations linear in the displacement
$\xi$ in the main approximation (\ref{eq:1}) of the spin-orbit
interaction. At $\bm{S}=0 \Rightarrow \bm{a}=0$ the equation
(\ref{eq:15}) is reduced to the geodesic deviation equation.},
   \begin{equation}\label{eq:15}
       \nabla_{u}\nabla_{u}\,\xi^{i} + E^{i}_{k}\,\xi^{k} = a^{i},
   \vspace{-2ex}
   \end{equation}
 where
   \[
   \nabla_{u}\nabla_{u} \,\bm{\xi} = \ddot{\bm{\xi}} +
    2\bm{\Omega}\times\dot{\bm{\xi}} +
    \dot{\bm{\Omega}}\times\bm{\xi} +
    \bm{\Omega}\times(\bm{\Omega}\times\bm{\xi}).
   \]
The dot designates the derivative with respect to pro\-per time
$\tau$. In the post-Newtonian approximation the spin-orbit force
applied to the intrinsic mass center of the rotating body is
   \begin{equation}\label{eq:16}
                ma^{i} = -c^{-1}B^{i}_{k}\,S^{k}.
   \end{equation}
This formula can be obtained, for example, by the matched
asymptotic expansions method \cite{8} or directly from the
Papapetrou equations with the supplementary conditions of Pirani
or Dixon (see \cite {16}; the distinctions between the exact
conditions of Pirani and Dixon, ibid.).

   In the equation (\ref{eq:15}) $E_{ik}$ is measured on the
fiducial geodesic $u$, but $B_{ik}$ in (\ref{eq:16}) ought to be
calculated in the frame comoving with the gyroscope mass center.
This "mixing" is admissible in the approximation linear in $S$
(\ref{eq:1}) and linear in $\xi$ (\ref{eq:15}) if the displacement
$\xi$ is induced by the spin-orbit interaction: $\xi \propto S,\;
\xi S \propto S^{2} \propto \xi^{2}= 0$. On the same ground we
transport the spin vector according to Fermi-Walker along the
fiducial geodesic,
   \begin{equation}\label{eq:17}
   \nabla_{u} \bm{S} = \dot{\bm{S}} + \bm{\Omega}\times\bm{S} = 0;
   \end{equation}
   \vspace{-3ex}
 \[ \dot{S}_{1} =  \omega_{s}S_{2},\;
    \dot{S}_{2} = -\omega_{s}S_{1},\;
    \dot{S}_{3} = 0.
 \]
The parallel transport equation (\ref{eq:17}) describes the known
geodetic precession (\ref{eq:4}):
   \begin{equation}\label {eq:18}
       S_{1} =  S\cos(\omega_{s}\tau + \beta),\;
       S_{2} = -S\sin(\omega_{s}\tau + \beta).
   \end{equation}
For the spin in the fiducial plane ($S_{3} = 0$), the equations of
the mass center motion (\ref{eq:15}), (\ref{eq:16}) are:
  \begin{equation}\label{eq:19}
  \left.
  \begin{array}{lll}
   \ddot{\xi}_{1}-2\omega_{s}\dot{\xi}_{2}&\ns+\ns&(E_{11}
   - \omega_{s}\,^{2})\xi_{1} =0, \\
   \ddot{\xi}_{2}+2\omega_{s}\dot{\xi}_{1}&\ns=\ns&0,
  \end{array}
  \right\}
  \end{equation}
  \vspace{-2ex}
  \begin{equation}\label{eq:20}
\ddot{\xi}_{3} + E_{33}\xi_{3} = 3g\epsilon
\frac{S_{1}}{\lambda}\,.
  \end{equation}

   The equations(\ref{eq:19}) describes the free oscillation
with the frequency $\omega'= \sqrt{E_{11} - 3\omega_{s}\,^{2}} =
n(1 - 3\epsilon/2)$, induced by an initial perturbation in the
fiducial plane. The difference of $\omega'$ from the orbital
frequency $\omega$ is caused by the general relativistic
pericenter drift of the perturbed quasielliptic orbit, $\omega -
\omega' = 3\epsilon n$. If the initial perturbation in the
fiducial plane is zero, the trajectory of gyroscope projection
onto the plane coincides with the circular geodesic.

   The equation of forced oscillation (\ref{eq:20}) along the orbital
normal
   \begin{equation}\label{eq:21}
        \ddot{\xi}_{3} + \omega^{2}\xi_{3} =
        3\epsilon\frac{S}{\lambda}g\cos(\omega_{s} + \beta)
   \end{equation}
proves to be quasiresonant due to proximity of the frequencies of
natural tidal oscillation $\sqrt{E_{33}} = \omega$ and of the
compelling force $\omega_{s}$. The difference of the frequencies
$\Delta \omega$ (\ref{eq:4}) preventing the oscillation from
resonance is equal to the geodetic precession $\Omega^{G}$. The
general solution of the equation (\ref{eq:21})
   \begin{equation}\label{eq:22}
                 \xi_{3} = A\cos\zeta - C\cos\eta,\;
                   \zeta = \omega_{s}\tau + \beta,\;
                    \eta = \omega\tau + \alpha
   \end{equation}
contains the amplitude $A$ (\ref{eq:5}) and two integration
constants, $C$ and $\alpha$. If $C = 0$, the oscillation
(\ref{eq:22}) describes the precession of the gyroscope orbit,
inclined by the angle $A/r = S/\lambda$ relative to the fiducial
plane, with the angular velocity of the geodetic precession
(\ref{eq:4})(Fig. 1 (a)). The evolution of the gyroscope orbital
momentum with arbitrary $C$ is presented on the Fig. 1 (b).
   \begin{figure}[h]
    \centering
    \includegraphics[width=0.4\textwidth]{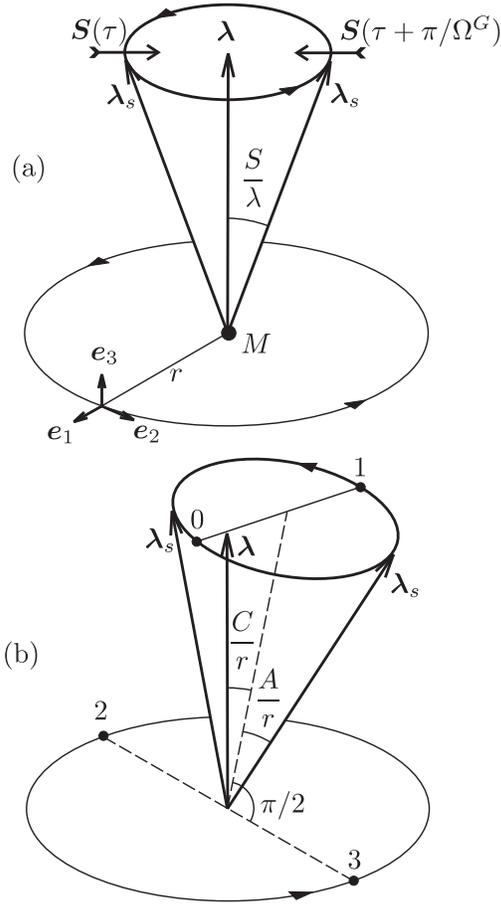}
    \caption{Orbit of the gyroscope. Orbital momenta of the fiducial
geodesic and the gyroscope are $\bm{\lambda}$ and
$\bm{\lambda}_{s}$ respectively. (a) Precession of the gyroscope
orbit at $C = 0$. (b) Variable inclination of the gyroscope orbit,
the constant $C$ is arbitrary. The orbital momentum
$\bm{\lambda}_{s}$ gets in the marked positions 0 and 1 when the
value $\sin((\eta - \zeta)/2)$ equals 0 and 1 respectively. In the
points 2 and 3 it turns out $\cos\eta=0$.}
   \end{figure}
 If $C = A$, the pure beating occurs,
   \begin{equation}\label{eq:23}
         \xi_{3} = 2A\sin\frac{\eta - \zeta}{2}
                     \sin\frac{\eta + \zeta}{2}.
   \end{equation}
The mass center makes the oscillation along the orbital normal
with the variable amplitude modulated by the geodetic precession
(\ref{eq:4}). The initial condition $\xi_{3}(\tau = 0) = 0$ is
provided by the choice of the constant $\alpha = -\beta$,
   \begin{equation}\label{eq:24}
                    \xi_{3} = 2A\sin\frac{\Delta \omega}{2}\tau\,
     \sin\left(\frac{\omega + \omega_{s}}{2}\tau + \beta\right)
   \end{equation}
and within a time $\tau \ll (\Delta\omega)^{-1}$ the oscillation
amplitude grows at a rate $A\,\Delta\omega$ (\ref{eq:6}),
(\ref{eq:7}). The condition $\dot{\xi}_{3}(\tau = 0) = 0$ fixes
the initial spin orientation $\sin\beta = 0$ along the radial
direction (see (\ref{eq:18})).

   The problem of measuring the oscillation (\ref{eq:24}) is
complicated by the circumstance that the initial perturbations
lead to natural tidal oscillation with the orbital frequency
$\omega^{2} = E_{33}$ (see (\ref{eq:21})). Therefore the
gyroscopes with antiparallel spins must be manufactured to be
coaxial. In order that the Newtonian harmonic oscillation due to
instrumental error be smaller than the relativistic oscillation
induced by the spin-orbit interaction, the strong restrictions on
the initial perturbations $\xi_{3}(0),\, \dot{\xi}_{3}(0)$ are
needed:
   \begin{equation}
          \xi_{3}(0) \ll \xi \sim A\,\Delta\omega\,\tau_{f},\quad
            \dot{\xi}_{3}(0) \ll \omega\xi,
   \end{equation}
where $\tau_{f}$ is the time of forming of the amplitude measured.

        \section{Effect of field oblateness}

   The Newtonian oblateness of the source does not lead to the
gyroscope forced oscillation. The oblateness affects the natural
tidal oscillation frequency $(\tl{E}_{33})^{1/2}$, orbital
frequency $\tl{\omega}$ and, consequently, the angular velocity
$\tl{\omega}_{s}$ of the spin rotation relative to orbital triad.
The two frequencies, $(\tilde{E}_{33})^{1/2}$ and
$\tl{\omega}_{s}$, enter the equation of motion of the gyroscope
mass center:
    \begin{equation}
               \ddot{\tl{\xi}}_{3} + \tl{E}_{33}\tl{\xi}_{3} =
 3\epsilon\frac{S}{\lambda}g\cos(\tl{\omega}_{s}\tau+\beta).
    \end{equation}
Considering only the quadrupole moment $J_{2}$ (Earth's $J_{2}
\simeq1\cdot10^{-3}$),  for an equatorial orbit we get
   \begin{eqnarray}
    \sqrt{\tl{E}_{33}}&\ns=\ns&\omega\,
          \left(1 + \frac{9}{4}J_{2}\frac{R^{2}}{r^{2}}\right), \\
           \tl{\omega}&\ns=\ns&\omega\,
          \left(1 + \frac{3}{4}J_{2}\frac{R^{2}}{r^{2}}\right), \\
       \tl{\omega}_{s}&\ns=\ns&\omega_{s}
          \left(1 + \frac{3}{4}J_{2}\frac{R^{2}}{r^{2}}\right),
    \end{eqnarray}
where $R$ is the equatorial radius of the source. The frequency
$(\tl{E}_{33})^{1/2}$ differs from the orbital frequency
$\tl{\omega}$ because of the Newtonian quadrupole precession
$\Omega^{J}$ of orbital plane,
   \begin{equation}\label{eq:30}
            \tl{\omega} - \sqrt{\tl{E}_{33}} =
 -\frac{3}{2}\omega J_{2}\frac{R^{2}}{r^{2}} = \Omega^{J}.
   \end{equation}
The gyroscope axis does not undergo the additional Newtonian
precession, $\tl{\omega} - \tl{\omega}_{s} = \Omega^{G}$. As a
result of the difference (\ref{eq:30}), the small denominator
(\ref{eq:4}) is changed
   \begin{eqnarray}
\Delta \tl{\omega}&\ns=\ns& \sqrt{\tl{E}_{33}} - \tl{\omega}_{s} =
        \Omega^{G} - \Omega^{J} \cong -\Omega^{J}\nonumber \\
         \mbox{}&\ns\cong\ns&
   \Delta \omega\frac{J_{2}}{\epsilon}\frac{R^{2}}{r^{2}}
   \end{eqnarray}
as well as the modulation period $\tl{T} = 2\pi/\Delta\tl{\omega}$
and the amplitude (\ref{eq:5})
   \begin{equation}
       \tl{A} = - A\,\frac{\Omega^{G}}{\Omega^{J}}
= \frac{S}{\lambda}\frac{\epsilon}{J_{2}}\frac{r^{2}}{R^{2}}\,.
   \end{equation}
The gyroscope orbital momentum vector describes a conic surface
with the apex angle $2\tl{A}/r$ and the time period $\tl{T}$. The
quadrupole precession period $\tl{T}$ of a near-Earth orbit is 2
months. For the pure beating
   \begin{equation}
\tl{\xi}_{3} = 2\tl{A}\,\sin\frac{\Delta\tl{\omega}}{2} \tau\,
       \sin\left[\left(\tl{\omega}_{s} +
       \frac{\Delta\tl{\omega}}{2}\right)\tau + \beta\right]
   \end{equation}
within the timescale $\tau \ll \tl{T}$ the oscillation increases
just as in the case of a spherically symmetric field (\ref{eq:6})
   \begin{equation}
                \tl{A}\,\Delta\tl{\omega} = A\,\Delta\omega =
                \frac{3}{2}\,\epsilon\,\frac{S}{\lambda}\,v.
   \end{equation}
The maximum amplitude formed in time $\tl{T}/2$ on a near-Earth
orbit for the gyroscope ${S}/ \lambda \sim 10^{-9}$ (\ref{eq:7})
   \begin{equation}
                   \tl{A} \sim 10^{-7} \mbox{ cm}
   \end{equation}
is several orders as good as the present-day limit of measuring small
oscillations.

        \section{Conclusions}

   General relativistic quasiresonant spin-orbit interaction leads
to the oscillation of the gyroscope mass center relative to the
fiducial geodesic along the orbital normal. The beating amplitude
does not include the speed of light and equals the ratio of the
intrinsic momentum of the gyroscope to its orbital momentum. The
modulation frequency equals the angular velocity of the geodetic
precession. The oscillation represents the precession of the
gyroscope orbital momentum. Within an acceptable time the
oscillation amplitude reaches the values that are amenable to
being analyzed experimentally.

   Taking into account the source oblateness decre\-ases the beating
amplitude and increases the modulation frequency by the factor
that is equal to the ratio of the quadrupole precession velocity
to the geodetic precession velocity. The period of the quadrupole
precession turns out to be a quite sufficient time to form a
measurable amplitude of the oscillation. The tidal acceleration,
providing the quasiresonant character of the oscillation, imposes
strong restrictions on the initial perturbations to distinguishing
the relativistic spin-orbit oscillation on the background of the
Newtonian tidal oscillation.

\end{document}